\documentclass[prl, twocolumn,showpacs]{revtex4}
\usepackage{dsfont}
\usepackage{enumerate}
\usepackage{amssymb,amsmath}
\usepackage{graphicx}
\usepackage{color}
\usepackage{amsmath}
\usepackage{bbm}
\usepackage{footnote}

\begin{document}

%
\title{Reconstruction of the environmental correlation function from single emitter photon statistics: a non-Markovian approach}
\author{
Faina Shikerman~$^{(1)}$, Lawrence P. Horwitz~$^{(1,2,3)}$ and Avi Pe'er~$^{(1)}$
\\ {\it Department of Physics, Bar-Ilan University,
Israel, Ramat-Gan, 52900$~^{(1)}$}\\ {\it School of Physics,
Tel-Aviv University, Israel, Ramat-Aviv, 69978$~^{(2)}$}\\ {\it
Department of Physics, Ariel University Center of Samaria, Israel,
Ariel, 40700$~^{(3)}$}} 

\begin{abstract}
We consider the two-level system approximation of a single emitter
driven by a continuous laser pump and simultaneously coupled to the
electromagnetic vacuum and to a thermal reservoir beyond the
Markovian approximation. We discuss the connection between a
rigorous microscopic theory and the phenomenological spectral
diffusion approach, used to model the interaction of the emitter
with the thermal bath, and obtained analytic expressions relating
the thermal correlation function to the single emitter photon
statistics.
\end{abstract}

 \maketitle



\makeatletter
\newcommand{\rmnum}[1]{\romannumeral #1}
\newcommand{\Rmnum}[1]{\expandafter\@slowromancap\romannumeral #1@}
\makeatother

\definecolor{Faina}{rgb}{0.4,0.5,0.9}
\definecolor{Gold}{rgb}{0.8,0.23,0.25}
%
%

\section{$\rm \bf \Large  \Rmnum{1}$ Introduction}

Single Molecule Spectroscopy (SMS) is a powerful experimental tool
with widespread applications in nano-technology, nano-biology,
quantum communication and quantum computation
\cite{Hong,Knill,Shih,Bouwmeester}. Given the possibility to isolate
a molecule, a quantum dot, or an atom allows for investigation of
the quantum dynamics of the system at the microscopic level. The
simplest method of the investigation consists of exciting the single
emitter by a laser and analyzing the outcome radiation. The
statistics of the photons, spontaneously emitted by such a single
quantum emitter, depends on its interaction with the environment and
permits extracting information on the latter at the level of atomic
distance and time scales. Thus, SMS is a key for nano-technology
advancement \cite{vanDijk,Santori,Katz,Orrit}.

The theory of an open system, extensively developed over the last
thirty years, combines phenomenological and microscopic approaches,
where the environment is typically modeled as a thermal bath of
harmonic oscillators
\cite{Weiss,Scully,CT,Vega,DGS,TYu,Kubo,Mori,Zwanzigjcp,Zwanzigjst,Nakajima}.
An important result of the theory is the derivation of a reduced
Liouville equation, obtained by tracing out the irrelevant degrees
of freedom. The most general integro-differential form of this
equation shows that all microscopic details of the coupling to a
reservoir are unified within the memory kernel given by the
environmental correlation function $\beta_T(t)$ (defined below). In
the so-called Markovian limit, the environmental correlation
function may be considered as a $\delta$-function of time (in
comparison to the time-scales of the unperturbed particle,
determined by the inverse of the eigneenergies of the system), which
leads to an irreversible, pure semigroup evolution. In practice,
however, the Markovian dynamics is only an approximation, whose
validity depends on several factors such as the density of
environmental states, temperature and the explicit form of the
coupling \cite{Vega,DGS,TYu}. At very high temperatures the
inaccuracy of the Markovian approximation is negligible
\cite{CL,Diosi}. However, at low temperatures, fast changing or
vanishing density of environmental states, as known to occur in
Photonic Band Gap materials, the evolution of the open system
corresponds to the non-Markovian regime, where the revival effects
are enhanced
while dissipation effects are minimized \cite{Weiss,Scully,CT,Vega,DGS,TYu,PBG}.\\

Up to now, the theoretical investigations of SMS were mainly based
on a phenomenological approach assuming that the coupling to a
thermal bath may be imitated by a real noise $\eta_t$ artificially
added to the unperturbed frequencies of the emitter
\cite{ZB,PRL,JCP}. The properties of this noise, called a spectral
diffusion process, and especially its two-times correlation function
$\langle\eta_t\eta_\tau\rangle$, are meant to reflect all possible
effects of the interaction with the surroundings. Such an
intuitively pleasing phenomenology simplifies the analysis, however,
it leaves unclarified the connection of the model to the rigorous
microscopic theory and does not give access to the temperature
dependent parameters of the reservoir.\\

In this article we provide a quantitative connection between the
spectral diffusion model and the microscopic method, thus allowing
SMS techniques to provide detailed information on the environmental
dynamics beyond the Markovian limit. In section $\rm \Rmnum{2}$ we
discuss the theoretical limitations of the spectral diffusion
approach through the comparison to a microscopic theory. In section
$\rm \Rmnum{3}$, using a version of the Dyson method and results of
the reduced propagator approach \cite{Vega,DGS,TYu}, we show that
whenever the environmental evolution is not specified and
interaction with a driving laser field is excluded, the spectral
diffusion model is valid, while the correlation function
$\langle\eta_{t}\eta_\tau\rangle$ is proportional to the real part
of the environmental correlation function $\beta_T(t-\tau)$. In
section $\rm \Rmnum{3}$, where the coupling to the laser is
restored, we notice that the real obstacle to the rigorous analysis
beyond the Markovian limit within the phenomenological theory is the
independent rates of variations assumption, which may be overcome
using the microscopic methods of \cite{Vega,DGS,TYu}, as done in
section $\rm \Rmnum{5}$. Finally, in section $\rm \Rmnum{6}$ we
establish the analytic relations beyond the Markovian approximation
between the single emitter photon statistics and the thermal
environmental correlation function, using a generalization of the
generating function technique for the photon counting events.
Section $\rm \Rmnum{7}$ summarizes
the main steps and concludes the article.\\
%

\section{$\rm \bf \Large \Rmnum{2}$ Generalized Langevin equation}

In this section we review the theoretical limitations of the
spectral diffusion approach and compare the Liouville equations
obtained by the phenomenological and the microscopic approaches. For
concreteness, we consider the two-level system approximation of a
single emitter, which is embedded in a thermal environment. For our
current purpose we exclude the interaction of the emitter with a
monochromatic laser pump and the electromagnetic vacuum, which will
be taken into account later. Thus, the unperturbed Hamiltonian of
the particle is
\begin{equation}\hat{H}_S=\frac{1}{2}\omega_0\hat{\sigma}_z,\label{Hs}
\end{equation}
where $\hat{\sigma}_i$
denotes the Pauli matrices. 
Within the spectral diffusion approach 
the two-level system density operator $\hat{\rho}(t|\eta_t)$ is a function
of a real process $\eta_t$, and obeys a stochastic Liouville
equation \cite{PRL,JCP,ZB}
\begin{equation}
\frac{d}{dt}\hat{\rho}(t|\eta_t)=i\left[\hat{\rho}
(t|\eta_t),\left(\hat{H}_S+\frac{1}{2}\eta_t\hat{\sigma}_z\right)\right].\label{LE0}
\end{equation}
Rewriting Eq.~(\ref{LE0}) in a vector notation defined by
$\vec{\rho}=\left(\rho_{\rm ee},\rho_{\rm gg},\rho_{\rm
ge},\rho_{\rm eg}\right)^{\rm T}$, where the suffixes ${\rm e}$ and
${\rm g}$ refer to the excited and the ground states respectively,
we have
\begin{equation}
\frac{d}{dt}\vec{\rho}(t|\eta_t)=\left[\hat{{\cal
L}}+\eta_t\hat{\Upsilon}\right]\vec{\rho}(t|\eta_t),\label{LE}
\end{equation}
where the superoperator
\begin{equation}
\hat{{\cal
L}}=\left(\begin{array}{cccc}0&0&0&0\\
0&0&0&0\\
0&0&i\omega_0&0\\
0&0&0&-i\omega_0\end{array}\right)\label{L}
\end{equation}
represents the reversible dynamics
$i\left[\hat{\rho}(t|\eta_t),\hat{H}_S\right]$, and the superoperator $\hat{\Upsilon}$, reflecting the contribution of the random
part
$i\left[\hat{\rho}(t|\eta_t),\frac{1}{2}\eta_t\hat{\sigma}_z\right]$,
is given by
\begin{equation}
\hat{\Upsilon}=\left(\begin{array}{cccc}0&0&0&0\\
0&0&0&0\\
0&0&i&0\\
0&0&0&-i\end{array}\right).\\\label{Upsilon}
\end{equation}

Regarding $\eta_t$ as white, Eq.~(\ref{LE}) is a standard Langevin
equation describing evolution of a Brownian particle \cite{Risken}.
The well-known generalization of Eq.~(\ref{LE}) for the
non-Markovian case of colored noise has the form of
\cite{Kubo,Mori,Zwanzigjcp,Zwanzigjst,Nakajima}
\begin{equation}
\frac{d}{dt}\vec{\rho}(t|\eta_t)=\left[\hat{{\cal
L}}_0+\eta_t\hat{\Upsilon}\right]\vec{\rho}(t|\eta_t)+
\int_{t_0}^t\hat{\gamma}(t,\tau)\vec{\rho}(\tau|\eta_\tau)d\tau,\label{GLE}
\end{equation}
where the memory kernel $\hat{\gamma}(t,\tau)$ is proportional to
the correlation function $\langle \eta_t\eta_\tau\rangle$, as a
manifestation of the fluctuation-dissipation theorem.
Eq.~(\ref{GLE}) shows that the dynamics of the open system generally
depends on its earlier states, because the environment is capable of
``remembering" the history of the particle's evolution. Since
$\eta_t$ and
$\hat{\gamma}(t,\tau)\propto\langle\eta_t\eta_{\tau}\rangle$ are
related by the fluctuation-dissipation theorem, it is inconsistent
to manipulate the former without altering the latter. Note that
although in the Markovian limit, $\langle
\eta_t\eta_\tau\rangle\propto\delta(t-\tau)$, Eq.~(\ref{GLE})
becomes local in time, Eqs.~(\ref{LE}-\ref{Upsilon}) cannot be
attributed to the Markovian limit of Eq.~(\ref{GLE}) because the
latter generically includes dissipation arising from the last term,
which is not reflected in Eqs.~(\ref{LE}-\ref{Upsilon}). Thus, for a
rigorous treatment of an open system interacting with the thermal
reservoir beyond the Markovian approximation, a priori, we cannot
use Eq.~(\ref{LE}) with $\eta_t$
colored, but must start the analysis from Eq.~(\ref{GLE}).\\

The phenomenological spectral diffusion approach was initiated
because the exact Hamiltonian of the entire system is often unknown,
making unavailable the projection of the time evolution equation of
the total system on the particle subspace. Yet, many systems may be
satisfactorily described by a simplified model Hamiltonian. A
judicious choice of such a Hamiltonian is capable not only of
providing the microscopic analogs of Eqs.~(\ref{LE},\ref{GLE}), but
also of shedding light on the microscopic origin of the spectral
diffusion parameters. For the particle $\otimes$ reservoir system it
is reasonable to choose the Lindblad Hamiltonian \cite{Weiss}
$$
\hat{H}=\left[\hat{H}_S+\hat{H}_B\right]+\hat{H}_{int}=$$\vspace{-.5cm}\begin{equation}=
\left[\hat{H}_0+\sum_\lambda\omega_\lambda \hat{b}^\dag_\lambda
\hat{b}_\lambda\right]+\sum_\lambda g_\lambda\left(\hat{L}^\dag
\hat{b}_\lambda+\hat{L}\hat{b}^\dag_\lambda\right), \label{LH}
\end{equation}
where $\hat{L}, \hat{L}^\dag$ are the particle operators in the
Schr\"{o}dinger picture, $\hat{b}_\lambda,\hat{b}^\dag_\lambda$ are
the annihilation and creation operators of bosons in the reservoir
mode $\lambda$, and $g_\lambda$ are the coupling coefficients.
Projecting the propagator of Eq.~(\ref{LH}) on fixed initial and
final environmental states, using the so-called reduced propagator
approach \cite{Vega,DGS,TYu}, within the second order approximation
in the coupling strength $g_\lambda$, yields the following master
equation for the particle:
%
%
$$
\!\!\!\!\frac{d}{dt}\hat{\rho}(t|z_{t})=i\left[\hat{\rho}(t|z_{t}),\hat{H}_S\right]+z_{t}^*\hat{L}\hat{\rho}
(t|z_{t})-
$$\vspace{-.5cm}$$-z_{t_0}\hat{L}^\dag\hat{\rho}(t|z_{t})+z_{t}\hat{\rho}(t|z_{t})\hat{L}^\dag-z_{t_0}^*\hat{\rho}
(t|z_{t})\hat{L}-$$\vspace{-.5cm}$$
\!\!\!\!\!-\!\int_{t_0}^td\tau\beta_T(\!t\!-\!\tau\!)\hat{L}^\dag{\rm
e}^{-i\hat{H}_S(t-\tau)}\hat{L}\hat{\rho}(\tau|z_{\tau}){\rm
e}^{i\hat{H}_S(t-\tau)}
-$$\vspace{-.5cm}\begin{equation}\-\!\int_{t_0}^td\tau\beta_T^*(\!t\!-\!\tau\!){\rm
e}^{-i\hat{H}_S(t-\tau)}\hat{\rho}(\tau|z_{\tau})\hat{L}^\dag{\rm
e}^{i\hat{H}_S(t-\tau)}\hat{L}.\label{dCrhoSB}
\end{equation}
Here, $z_{t}$ 
is a complex function representing a time dependent average of the environmental states, weighted by $g_\lambda$ \cite{Vega,DGS,TYu} (see also
Appendix B), and
\begin{equation}
\beta_T(t)\!=\!\int_0^\infty\!\!\!\!d\omega
J(\omega)\!\left[\!\coth\!\left(\!\frac{\omega}{2\kappa_{\rm B}
T}\!\right)\!\cos(\omega t) \!-\!i\sin(\omega t)\right] \label{beta}
\end{equation}
(in the continuum
limit $\lambda\rightarrow\omega$) is a transform of the environmental spectral function
\begin{equation}
J(\omega)= g(\omega)^2D(\omega),\label{3JCL}
\end{equation}
where $D(\omega)$ is the density of the environmental states. Integrating over the initial and the final reservoir states
one can show that \cite{Vega,DGS,TYu}
\begin{equation}
\langle z_{t}\rangle=0, ~~ \langle z_{t} z_{\tau}\rangle=0, ~~
\langle z_{t} z^*_\tau\rangle=\beta_T(t-\tau),
\label{zmoment}
\end{equation}
implying that $z_{t}$ may be interpreted as a random zero mean
process with a correlation function $\beta_T(t)$ (Eq.~(\ref{beta})).
Hence, the reduced density matrix $\hat{\rho}(t|z_{t})$ in
Eq.~(\ref{dCrhoSB}) is conditioned by a given environmental
evolution path, analogous to the phenomenological density matrix
$\hat{\rho}(t|\eta_t)$ in Eqs.~(\ref{LE0},\ref{GLE}), conditioned by
a realization of the spectral diffusion process $\eta_t$. In other
words, Eq.~(\ref{dCrhoSB}) is the microscopic counterpart of
Eq.~(\ref{GLE}).\\

To demonstrate a disagreement between the spectral diffusion method
Eqs.~(\ref{LE}-\ref{Upsilon}) and the microscopic approach
Eq.~(\ref{dCrhoSB}) we shall specify the coupling operator
$\hat{L}$. Since the interaction of the emitter with a thermal
environment is usually insufficient to generate transitions between
the states of an unperturbed system, it is manifested as a random
noise perturbing only the energy levels. This effect accounts for a
self adjoint coupling $\hat{L}=\hat{L}^\dag=\hat{K}$, where for a
two-level system $\hat{K}=\hat{\sigma}_z$. With such a substitution,
well-known as the spin-boson model \cite{Weiss}, rewriting
Eq.~(\ref{dCrhoSB}) in vector notation leads to a Liouville equation
in the form of Eq.~(\ref{GLE}), where $\eta_t\Rightarrow \Delta_{
z_{t}}\equiv(z_{t}^*-z_{t_0})$, $\hat{{\cal L}}_0=\hat{{\cal L}}$ is
given by Eq.~(\ref{L}),
\begin{equation}
\hat{\Upsilon}=2\left(\!\begin{array}{cccc}{\rm Re}[\Delta_{z_{t}}]\!\!&0\!\!&0\!\!&0\!\!\\
0\!\!\!&-{\rm Re}[\Delta_{z_{t}}]\!\!&0\!\!&0\!\!\\
0\!\!&0\!\!\!&-i{\rm Im}[\Delta_{z_{t}}]\!\!&0\!\!\\
0\!\!&0\!\!&0\!\!\!&i{\rm
Im}[\Delta_{z_{t}}]\!\!\end{array}\right),\label{Ups}
\end{equation}
and the memory kernel $\hat{\gamma}(t)$ is given by
\begin{equation}
\hat{\gamma}(t)=-2{\rm Re}[\beta_T(t)]\left(\begin{array}{cccc}1&0&0&0\\
0&1&0&0\\
0&0&{\rm e}^{i\omega_0 t}&0\\
0&0&0&{\rm e}^{-i\omega_0
t}\end{array}\right).\\\label{4gamma}
\end{equation}
%
Superposing Eqs.~(\ref{LE}-\ref{Upsilon}) and
Eqs.~(\ref{GLE},\ref{L},\ref{Ups},\ref{4gamma}) in the Markovian
limit, we see that the only way to make the phenomenological and the
microscopic methods agree is by restricting $ \Delta_{ z_{t}}$ and
$\beta_T(t-\tau)$ to be purely imaginary. Inspecting
Eq.~(\ref{beta}) we see that the latter constraint may be only
approximately satisfied for $T\rightarrow 0$, which corresponds to
the non-Markovian regime. 

\section{$\rm \bf \Large  \Rmnum{3}$ Dyson equation}

In this section we show that despite the discussed limitations of
the phenomenological approach described above; the spectral
diffusion model Eq.~(\ref{LE}), with $\eta_t$ white or colored, may
be justified under standard experimental conditions, while $\langle
\eta_t \eta_\tau\rangle$ may be identified, up to a constant, with
the real part of the microscopic environmental correlation function
$\beta_T(t-\tau)$. Note that in practice the evolution of the
environment is usually not determined on the microscopic level, so
that the quantity measured is not $\hat{\rho}(t|\eta_t)$ [or
$\hat{\rho}(t|z_{t})$], but rather its mean value, given as
\begin{equation}
\hat{\rho}_S(t)=\langle \hat{\rho}(t|\eta_t)\rangle ~~~ [{\rm or}
~\hat{\rho}_S(t)=\langle \hat{\rho}(t|z_{t})\rangle],\label{mrho}
\end{equation}
which is the averaged reduced density matrix standardly used in the
literature. For a general stochastic equation in the form of
\begin{equation}
\frac{d}{dt}\vec{\rho}(t|\eta_t)=\hat{\cal O}\left[\vec{\rho}(t|\eta_t),t\right]+\eta_t\hat{\Upsilon}\vec{\rho}(t|\eta_t),\label{SchEq}
\end{equation}
where $\hat{\cal O}$ is any functional of $\vec{\rho}(t|\eta_t)$,
the equation of motion for the expectation value
$\vec{\rho}_S(t)=\langle \vec{\rho}(t|\eta_t)\rangle$ may be
obtained using an analogue of the Dyson method. We expand the
propagator of Eq.~(\ref{SchEq}) by iteration as
\begin{equation}
\hat{U}(t,t_0|\eta_t)=\sum_{n=0}^\infty \hat{U}^{(n)}(t,t_0|\eta_t),
\label{AGLExp}
\end{equation}
where
$$
\hat{U}^{(n)}(t,t_0|\eta_t)\!=\!\int_{t_0}^t
dt_n\int_{t_0}^{t_n}dt_{n-1}\!_{\cdots}\!\int_{t_0}^{t_2}dt_1\times$$\vspace{-.5cm}\begin{equation}\times
 \hat{U}_0(t,t_n) \eta_{t_n}\hat{\Upsilon} \hat{U}_0(t_n,t_{n-1})\eta_{t_{n-1}}\hat{\Upsilon}
 _{\cdots}
\eta_{t_1}\hat{\Upsilon} \hat{U}_0(t_1,t_0),\label{AUn}
\end{equation}
and $\hat{U}_0(t,t')$ obeys Eq.~(\ref{SchEq}) with $\eta_t=0$, i.e.,
\begin{equation}
\frac{d}{dt}\hat{U}_0(t,t')=\hat{\cal O}\left[\hat{U}_0(t,t'),t\right].\label{GL0E}
\end{equation}
Each summand in the rhs of Eq.~(\ref{AGLExp}) suggests the
well-known interpretation of the ``free" evolution, generated by the
unperturbed propagator $\hat{U}_0(t,t')$, interrupted $n=0,1,2,...$
times by the random ``potential" $\eta_t\hat{\Upsilon}$. Acting with the
average of Eq.~(\ref{AGLExp}) on the initial state vector $\vec{\rho}(t_0)$ yields an expression for $\vec{\rho}_S(t)$ in terms of a sum of
integrals including the zeroth, first, second, etc. moments of $\eta_t$, as shown in Appendix A.\\

 Further progress is possible if $\langle\eta_{t_n}\eta_{t_{n-1}}...\eta_{t_1}\rangle$ may be factorized
into a product of correlation functions of a lower order, as, for
example, occurs for a Gaussian noise. Due to the special properties
of the latter, assuming it is of zero mean, all the odd moments
vanish, while every $\hat{U}^{(n)}(t,t_0|\eta_t)$ defined by
Eq.~(\ref{AUn}) with $n$ even gives rise to
$n!/\left(2^{\frac{n}{2}}\frac{n}{2}!\right)$ terms, differing one
from another only in the way $\langle \eta_{t_n}\eta_{t_{n-1}}\cdots
\eta_{t_1}\rangle$ is factorized \cite{Risken}. These terms may be
represented by Feynman diagrams and classified according to their
physical meaning. For example,
$$\hat{U}^{(4)}(t,t_0|\eta_t)=\!\int_{t_0}^t \!\!d_{t_4}\!\int_{t_0}^{t_4}\!\! d_{t_3}\!
\int_{t_0}^{t_3}\!\! d_{t_2}\!\int_{t_0}^{t_2} \!\!d_{t_1}\langle \eta_{t_4}\eta_{t_3} \eta_{t_2}\eta_{t_1}\rangle\times
$$\vspace{-.4cm}
\begin{equation}
\times \hat{U}_0(t,t_4)\hat{\Upsilon} \hat{U}_0(t_4,t_3)\hat{\Upsilon}\hat{U}_0(t_3,t_2)\hat{\Upsilon} \hat{U}_0(t_2,t_1)\hat{\Upsilon}
\hat{U}_0(t_1,t_0)\label{U4}\end{equation}
leads to three different
diagrams resulting from
$$
\langle \eta_{t_4}\eta_{t_3} \eta_{t_2}\eta_{t_1}\rangle= \langle
\eta_{t_4}\eta_{t_3}\rangle \langle \eta_{t_2}\eta_{t_1}\rangle+$$\vspace{-.5cm}
\begin{equation}+
\langle \eta_{t_4}\eta_{t_1}\rangle \langle
\eta_{t_3}\eta_{t_2}\rangle+ \langle \eta_{t_4}\eta_{t_2}\rangle
\langle \eta_{t_3}\eta_{t_1}\rangle.\label{factor}
\end{equation}
Now, if the magnitude of the correlation function $\langle
\eta_{t}\eta_{\tau}\rangle$ is likely to decrease with an increase
of $t-\tau$, which can be true even in the non-Markovian regime, it
is clear that the first summand $\langle \eta_{t_4}\eta_{t_3}\rangle
\langle \eta_{t_2}\eta_{t_1}\rangle$ on the rhs of
Eq.~(\ref{factor}) gives rise to a diagram which dominates over the
other two diagrams arising from $\langle\eta_{t_4}\eta_{t_1}\rangle
\langle \eta_{t_3}\eta_{t_2}\rangle$ and $\langle
\eta_{t_4}\eta_{t_2}\rangle \langle \eta_{t_3}\eta_{t_1}\rangle$,
because the time interval between the subsequent moments is always
the smallest. 
Hence, to a reasonable approximation, all the diagrams including the
correlation functions of a pair of non subsequent times may be
neglected, which leads to a master equation
\begin{equation}
\frac{d}{dt}\vec{\rho}_S(t)=\hat{\cal O}\left[\vec{\rho}_S(t),t\right]+ \int_{t_0}^{t}
\hat{{\cal M}}(t,\tau)\vec{\rho}_S(\tau)d\tau,
\label{GDE}
\end{equation}
whose memory kernel is given by
\begin{equation}
\hat{{\cal M}}(t,\tau)=\langle \eta_t\eta_\tau\rangle\hat{\Upsilon}
\hat{U}_0(t,\tau)\hat{\Upsilon}.\label{calM}\end{equation}
This quite general result means that the effect of a colored noise,
arising from an interaction with an environment and rigorously
described by Eq.~(\ref{GLE}) would be indistinguishable from that
predicted by Eq.~(\ref{LE}), provided the latter is constructed such
that
the resulting Dyson equations (\ref{GDE}) coincide.\\

To apply Eqs.~(\ref{GDE},\ref{calM}) to
the spectral diffusion model
Eqs.~(\ref{LE}-\ref{Upsilon}), we set $\hat{\cal O}\left[\vec{\rho}(t|\eta_t),t\right]=\hat{\cal L}\vec{\rho}(t|\eta_t)$, where $\hat{{\cal L}}$ is given by Eq.~(\ref{L}).
In such a case the
solution for $\hat{U}_0(t,t')$ becomes trivial and yields
\begin{equation}
\hat{{\cal M}}(t,\tau)\!=-\langle \eta_t\eta_\tau\rangle\left(\!\begin{array}{cccc}0&0&0&0\\
0&0&0&0\\
0&0&{\rm e}^{i\omega_0(t-\tau)}&0\\
0&0&0&{\rm
e}^{-i\omega_0(t-\tau)}\end{array}\!\!\!\right).\label{M}
\end{equation}
On the other hand, setting $\hat{L}=\hat{L}^\dag=\hat{K}$ and
assuming the reservoir is initially prepared in Boltzmann
equilibrium at temperature $T$, eliminating the environmental
degrees of freedom in Eq.~(\ref{dCrhoSB}) leads to
\cite{Vega,DGS,TYu}
\begin{equation}\begin{array}{c}
\frac{d}{dt}\hat{\rho}_S(t)=-i\left[\hat{H}_S,\hat{\rho}_S(t)\right]
-\\\\
\int_{t_0}^t\beta_T(t-\tau)\left[\hat{K}, {\rm e}^{-i\hat{H}_S(t-\tau)}\hat{K}\hat{\rho}_S(\tau){\rm
e}^{i\hat{H}_S(t-\tau)}\right]d\tau-\\\\
- \int_{t_0}^t\beta_T^*(t-\tau)\left[{\rm
e}^{-i\hat{H}_S(t-\tau)}\hat{\rho}_S(\tau)\hat{K}{\rm e}^{i\hat{H}_S(t-\tau)},\hat{K} \right]d\tau,\end{array}
\label{dtroalphaT}
\end{equation}
%
%
%
Substituting $\hat{H}_S=\frac{1}{2}\omega_0\hat{\sigma}_z$, $\hat{K}=\hat{\sigma}_z$ and
rewriting Eq.~(\ref{dtroalphaT}) in vector notation we obtain an equation in the form of
Eq.~(\ref{GDE}), where $\hat{\cal O}\left[\vec{\rho}(t|\eta_t),t\right]=\hat{\cal L}\vec{\rho}(t|\eta_t)$,
with $\hat{{\cal L}}$ given by Eq.~(\ref{L}), and
\begin{equation}
\!{\cal M}_T(\!t\!-\!\tau\!)\!=\!-4{\rm
Re}\!\left[\beta_T(\!t\!-\!\tau\!)\right]\!\!
\left(\!\begin{array}{cccc}0&0&0&0\\
0&0&0&0\\
0&0&\!\!{\rm e}^{i\omega_0(t-\tau)}&0\\
0&0&0&\!\!\!\!\!\!\!{\rm
e}^{-i\omega_0(t-\tau)}\end{array}\!\!\!\right),\label{Mmic}
\end{equation}
which constitutes the microscopic analog of the phenomenological
memory kernel Eq.~(\ref{M}). We see that in case of a stationary
process$^{1}$\footnotetext[1]{The model
Eq.~(\ref{LH}) is sufficient to yield only a stationary environmental
noise, whose memory kernel $\beta_T(t-\tau)$ Eq.~(\ref{beta}) is
invariant under time translation. An extension of the method to
non-stationary noises may be achieved by a ``nested doll" model,
where the particle $\otimes$ reservoir system is
itself considered as a subsystem of a larger environment.} the two
coincide, provided
\begin{equation}
4{\rm Re} \left[\beta_T(t-\tau)\right]=
\langle\eta_t\eta_\tau\rangle.\end{equation}
In such a way, under all mentioned approximations, the
phenomenological model Eqs.~(\ref{LE}-\ref{Upsilon}), driven by a
random real spectral diffusion process $\eta_t$, white or colored,
leads to the same marginal master equation for the reduced density
matrix $\hat{\rho}_S(t)$ as the microscopic approach. Therefore,
despite the limitations discussed in the previous section, it may
serve as a shorter effective form for the description of the
two-level system conditional probability density matrix dynamics,
whenever the environmental evolution cannot be fixed on the
microscopic level. In other words, the spectral diffusion
correlation function $\langle\eta_t\eta_\tau\rangle$ may be assumed
to be some arbitrary function of time (i.e., not necessarily a
$\delta$-function), as it may be rigorously identified with the real
part of the environmental correlation function, i.e., $4{\rm
Re}\left[\beta_T(t-\tau)\right]$.

%

\section{$\rm \bf \Large \Rmnum{4}$ Independent rates of variation approximation}

Under the assumptions discussed in the previous section we saw that
when the environmental evolution is not determined, the spectral
diffusion model
yields a master equation matching the one obtained by the exact
microscopic analysis. Thus, the inconsistency of the
phenomenological equation (\ref{LE}) with the
fluctuation-dissipation theorem in case of colored noise is
effectively eliminated after averaging over all the realizations of
$\eta_t$, and the Dyson equation (\ref{GDE}) is then valid also
beyond the Markovian limit. Let us, however, recall that such a
conclusion has been obtained assuming
$\hat{H}_S\propto\hat{L}=\hat{L}^\dag$, i.e., excluding single
emitter interaction with the driving laser field. Restoring the
laser pump within the rotating wave approximation \cite{CT}, the
two-level system Hamiltonian is no longer diagonal in the eigenbasis
of $\hat{H}_S$ introduced by Eq.~(\ref{Hs}), and is given by
\begin{equation}
\hat{H}_S(t)=\frac{1}{2}\omega_0\hat{\sigma}_z
+\Omega_0\cos(\omega_Lt)\left[\hat{\sigma}_++\hat{\sigma}_-\right],\label{HtS}
\end{equation}
where $\omega_L$ and $\Omega_0=-\frac{{\rm d}_{\rm eg}{\cal
E}}{\hbar}$ are the angular and the Rabi frequencies of the laser
(${\rm d}_{\rm eg}={\rm d}_{\rm ge}$ are the matrix elements of the
off-diagonal electric dipole moment, and ${\cal E}$ is the laser
amplitude). Switching to a rotating frame by the unitary
transformation
$\hat{R}(t)=\exp\left[i\frac{\omega_L}{2}t\hat{\sigma}_z \right]$
\cite{JCP}, allows eliminating the explicit time dependence, and
yields
\begin{equation}
\hat{H}_S=\frac{1}{2}\Delta_L\hat{\sigma}_z
+\frac{1}{2}\Omega_0(\hat{\sigma}_++\hat{\sigma}_-),\label{HS}
\end{equation}
where $\Delta_L\equiv\omega_0-\omega_L$ is the detuning, which entails the corresponding Liouville operator
\begin{equation}
\hat{{\cal L}}=\left(\begin{array}{cccc}0&0&-i\frac{\Omega_0}{2}&i\frac{\Omega_0}{2}\\
0&0&i\frac{\Omega_0}{2}&-i\frac{\Omega_0}{2}\\
-i\frac{\Omega_0}{2}&i\frac{\Omega_0}{2}&i\Delta_L&0\\
i\frac{\Omega_0}{2}&-i\frac{\Omega_0}{2}&0&-i\Delta_L\end{array}\right).\label{L1}
\end{equation}
According to the spectral diffusion approach, the master equation for
the particle is now obtained by Eq.~(\ref{LE}) where ${\hat{\cal L}}$ is given by Eq.~(\ref{L1}),
while the
interaction with the thermal environment, described by $\eta_t$ and
$\hat{\Upsilon}$ Eq.~(\ref{Upsilon}), stays unaltered \cite{PRL,JCP,ZB}.
It is evident that 
by simple addition of a real random process to the unperturbed
emitter transition frequency $\omega_0$, the phenomenological model
independently imposes the rates of variation associated with the
interaction with the thermal bath and the laser, as if each coupling
acted alone. In other words, the spectral diffusion approach is
based on the so-called ``independent rates of variation"
approximation, neglecting the effect of possible correlation between
the laser and the thermal reservoir, induced by the coupling to the
two-level system. Generally, this approximation is legitimate when
the time scale of the particle evolution induced by the coupling to
the laser, i.e. $\Omega_0^{-1}$, is much longer than the correlation
time $\tau_c$ of the relaxation process (determined by the decay of
$\langle\eta_t\eta_\tau\rangle$), and becomes strictly exact only if
$\tau_c\rightarrow 0$ \cite{CT}. This statement is supported by the
microscopic analysis, since considering the terms ${\rm
e}^{-i\hat{H}_S(t-\tau)}\hat{L}\hat{\rho}(\tau|z_{\tau}){\rm
e}^{i\hat{H}_S(t-\tau)}$ and ${\rm
e}^{-i\hat{H}_S(t-\tau)}\hat{\rho}(\tau|z_{\tau})\hat{L}^\dag{\rm
e}^{i\hat{H}_S(t-\tau)}$ in the integrands of Eq.~(\ref{dCrhoSB}),
it may be readily noted that an incompatibility of the particle
Hamiltonian $\hat{H}_S$ with the Lindblad operators
$\hat{L},\hat{L}^\dag$ has an effect only beyond the Markovian
limit. Therefore, the independent rates of variation assumption,
intrinsically inserted into the phenomenological spectral diffusion
method, constitutes an essential consequence of the Markovian
approximation, and it is natural to expect that for
$\left[\hat{H}_S,\hat{L}\right]\neq0$ the phenomenological Dyson
equation (\ref{GDE}) will no longer coincide with the analogous
equation (\ref{dtroalphaT}) obtained
by the microscopic analysis beyond the Markovian limit.\\

To see explicitly how the independent rates of variation approximation alters the results of
Section $\rm \Rmnum{3}$, we first revise the phenomenological scheme, outlined in Eqs.~(\ref{GDE}-\ref{calM}).
Setting  $\hat{\cal O}\left[\vec{\rho}(t|\eta_t),t\right]=\hat{\cal L}\vec{\rho}(t|\eta_t)$, where now $\hat{{\cal L}}$ is given by Eq.~(\ref{L1}),
we arrive at a master equation in the form of Eq.~(\ref{GDE}),
whose memory kernel is
\begin{equation}
\hat{\cal M}(t)\!=\left(\!\begin{array}{cccc}0&0&0&0\\
0&0&0&0\\
0&0&B(t)&C(t)\\
0&0&C(t)&B^*(t)\end{array}\!\!\!\right),\\\label{ML}
\end{equation}
where, using the definition of the generalized Rabi frequency
$\Omega\equiv\sqrt{\Delta^2_L+\Omega_0^2}$,
$$
B(t)=-\frac{\langle
\eta_t\eta_0\rangle}{2\Omega^2}\left[\Omega_0^2+\cos\left[\Omega
t\right]\left(
\Omega_0^2+2\Delta_L^2\right)+2i\Delta_L\Omega\sin\left[\Omega
t\right]\right],$$\vspace{-.5cm}\begin{equation} C(t)=\langle
\eta_t\eta_0\rangle\frac{\Omega_0^2}{\Omega^2}\sin^2\left[\frac{\Omega}{2}t\right].
\end{equation}
On the other hand,
substituting $\hat{H}_S$, given by Eq.~(\ref{HS}) into Eq.~(\ref{dtroalphaT}) (with $\hat{K}=\hat{\sigma}_z$),
we arrive at a master equation in the form of Eq.~(\ref{GDE}), where $\hat{\cal L}$, once again, is given by Eq.~(\ref{L1}),
whereas the memory kernel takes the form

\begin{equation}
\hat{\cal M}_T(t)=\left(\begin{array}{cccc}0&0&0&0\\
0&0&0&0\\
A_T(t)&A_T(t)&B_T(t)&C_T(t)\\
A_T^*(t)&A_T^*(t)&C_T(t)&B_T^*(t)\end{array}\right),\label{Mmic2}
\end{equation}
where
%
$$
A_T(t)=2i{\rm
Im}\left[\beta_T(t)\right]\frac{\Omega_0}{\Omega^2}\left\{\Delta_L\left(1-\cos\left[\Omega t\right]\right)
-i\Omega\sin\left[\Omega t\right]\right\},$$ \vspace{-.5cm}$$
\!\!\!B_T(t)\!=\!-\frac{2{\rm
Re}\left[\!\beta_T(t)\!\right]}{\Omega^2}\!\left\{\!\Omega_0^2\!+\!(2\Delta^2_L\!+\!\Omega_0^2)\!\cos\left[\Omega t\right]\!+\!
2i\Delta_L\Omega\sin\left[\Omega t\right]\!\right\},$$
\vspace{-.5cm}\begin{equation} C_T(t)=4{\rm
Re}\left[\beta_T(t)\right]\frac{\Omega_0^2}{\Omega^2}\sin^2\left[\frac{\Omega}{2}
t\right]. \label{TABC}
\end{equation}
Remembering that $\langle \eta_t\eta_0\rangle=4{\rm
Re}\left[\beta_T(t)\right]$, we compare Eqs.~(\ref{ML}-\ref{TABC})
and see: $B(t)=B_T(t)$, $C(t)=C_T(t)$, while the difference enters
through $A_T(t)$, which vanishes in the phenomenological case. Note
that $A_T(t)$, unlike $B_T(t)$ and $C_T(t)$, is proportional to
${\rm Im}\left[\beta_T(t)\right]$. In the Markovian limit the real
part (dissipation) of the environmental correlation function
dominates over the imaginary part (fluctuation). This means that
$A_T(t)$, inducing disagreement between the microscopic theory and
the spectral diffusion model, becomes important in the non-Markovian
regime. The Markovian limit of Eqs.~(\ref{ML},\ref{Mmic2}) may be
recovered by setting $t\!=\!0$, which yields
$$\begin{array}{c}
\hat{\cal M}(0)=\hat{\cal M}_T(0)=-4{\rm Re}\left[\beta_T(0)\right]\left(\begin{array}{cccc}0&0&0&0\\
0&0&0&0\\
0&0&1&0\\
0&0&0&1\end{array}\right).\end{array}$$
Hence, in the Markovian approximation the phenomenological method
gives a result which coincides with the microscopic approach, as
expected. In summary so far, we conclude that due to the independent
rates of variation assumption (and not because of the apparent
incompatibility with the fluctuation-dissipation theorem) the
spectral diffusion model Eq.~(\ref{LE}) does not consistently
describe the SMS experiments beyond the Markovian approximation. The
microscopic approach, on the other hand, permits avoiding the
independent rates of variation assumption, and also clarifies the
role of the environmental spectral function $J(\omega)$
Eq.~(\ref{3JCL}) and the temperature of the thermal reservoir $T$.

\section{$\rm \bf \Large \Rmnum{5}$ Master equation for a two-level system interacting
with two reservoirs}

In this section we set up a reduced master equation describing the
SMS experiments beyond the Markovian limit using the microscopic
methods of \cite{Vega,DGS,TYu}. Recall that we considered a
two-level system emitter continuously driven by a classical
monochromatic laser field. Were complete isolation from the
environment realistic, the particle would undergo simple Rabi
oscillations \cite{CT}. In practice, the unavoidable interaction
with the electromagnetic vacuum induces incoherent decay transitions
accompanied by events of the spontaneous photon emission, while the
coupling to the thermal environment is associated with the
origin of the spectral diffusion noise.\\

The total system, in the rotating waves approximation, is described by the
Hamiltonian (see Appendix B)
\begin{equation}
\begin{array}{c}\!\hat{H}\!= \!\frac{\Delta_L}{2}\hat{\sigma}_z\!+\!\frac{\Omega_0}{2}(\hat{\sigma}_+
\!+\!\hat{\sigma}_-)\!+\!
\sum_{\mu}(\omega_{\mu}\!-\!\omega_L)\hat{a}^{\dag}_{\mu}\hat{a}_{\mu}\!+\!\sum_{\lambda}\omega_{\lambda}\hat{b}^{\dag}_{\lambda}\hat{b}_{\lambda}\!+\\
\\+\sum_{\mu}p_{\mu}(\hat{\sigma}_+
\hat{a}_{\mu}+\hat{\sigma}_- \hat{a}^{\dag}_{\mu})+\sum_{\lambda}g_{
\lambda}\hat{\sigma}_z(\hat{b}^{\dag}_{\lambda}+\hat{b}_{\lambda}),\end{array}\label{Hrwa}
\end{equation}
where the first two terms describe the Hamiltonian of the two-level system driven by the laser; the 3-4th
terms are the free Hamiltonians of the vacuum and the thermal bath,
whose interaction with the particle is described by 5th and the 6th
terms respectively. The theory of the reduced propagator and its
application to the derivation of the marginal master equation was
comprehensively developed for a particle interacting with a single
bosonic field in \cite{Vega,DGS,TYu}. In case of a particle simultaneously
coupled to several baths, the derivation of the master
equation might be complicated by the entanglement arising between
the reservoirs from the induced indirect interaction. However, this
interaction, as shown in Appendix B, is manifested in terms
going beyond the second order in coupling strength, and may be
neglected if the interaction is weak. 
Assuming in our case that the two-level system is weakly coupled to
both reservoirs, it is justified to approximate the reduced
evolution of the particle imposing the contributions arising from
the interaction with each one of the reservoirs independently, which
yields
$$
\frac{d}{dt}\hat{\rho}_S(t)=i\left[\hat{\rho}_S(t),\hat{H}_S\right]-$$\vspace{-.5cm}$$-\int_{t_0}^t
d\tau\alpha(t-\tau)\left[\hat{\sigma}_+ ,{\rm
e}^{-i\hat{H}_S(t-\tau)}\hat{\sigma}_-\hat{\rho}_S(\tau){\rm e}^{i\hat{H}_S(t-\tau)}\right]-$$\vspace{-.5cm}
$$ \!\!\!\!-\int_{t_0}^t d\tau\alpha^*(t-\tau)\left[{\rm e}^{-i\hat{H}_S(t-\tau)}\hat{\rho}_S(\tau)
\hat{\sigma}_+{\rm e}^{i\hat{H}_S(t-\tau)},\hat{\sigma}_-\right]-$$
\vspace{-.5cm}$$
-\int_{t_0}^td\tau\beta_T(t-\tau)\left[\hat{\sigma}_z, {\rm
e}^{-i\hat{H}_S(t-\tau)}\hat{\sigma}_z\hat{\rho}_S(\tau){\rm e}^{i\hat{H}_S(t-\tau)}\right]-$$\vspace{-.5cm}
\begin{equation} -\int_{t_0}^t d\tau\beta_T^*(t-\tau)\left[{\rm e}^{-i\hat{H}_S(t-\tau)}\hat{\rho}_S(\tau)
\hat{\sigma}_z{\rm e}^{i\hat{H}_S(t-\tau)},\hat{\sigma}_z \right],
\label{dtroConv}
\end{equation}
where $\hat{H}_S$ is given by Eq.~(\ref{HS}). Taking the Markovian
limit of the electromagnetic vacuum correlation function
$\alpha(t-\tau)\propto\delta(t-\tau)$ and rewriting
Eq.~(\ref{dtroConv}) in vector notation, we find
\begin{equation}
\frac{d}{dt}\vec{\rho}_S(t)=\left[\hat{\cal
L}+\hat{\Gamma}\right]\vec{\rho}_S(t)+\int_{t_0}^t\hat{\cal M}_T(t-\tau)\vec{\rho}_S(\tau)
d\tau,\label{drhoNM}
\end{equation}
where
\begin{equation}
\hat{\cal
L}\!=\!\!\left(\!\!\begin{array}{cccc}-\gamma&0&-i\frac{\Omega_0}{2}&i\frac{\Omega_0}{2}\\
0&0&i\frac{\Omega_0}{2}&-i\frac{\Omega_0}{2}\\
-i\frac{\Omega_0}{2}&i\frac{\Omega_0}{2}&-\frac{\gamma}{2}+i\Delta_L&0\\
i\frac{\Omega_0}{2}&-i\frac{\Omega_0}{2}&0&-\frac{\gamma}{2}-i\Delta_L\end{array}\!\!\right),\label{L0}
\end{equation}
\begin{equation}
\hat{\Gamma}=\left(\begin{array}{cccc}0&0&0&0\\
\gamma&0&0&0\\
0&0&0&0\\
0&0&0&0\end{array}\right),\label{Gamma}
\end{equation}
$\gamma$ is the spontaneous emission rate \cite{CT,Scully}, and the
thermal memory kernel $\hat{\cal M}_T(t)$, already calculated, is
given by Eqs.~(\ref{Mmic2},\ref{TABC}). Eq.~(\ref{drhoNM}) is the
desired reduced master equation, unrestricted by the assumption of
independent rates of variations, for a driven two-level system
interacting with two bosonic reservoirs: one within the Markovian
approximation (the electromagnetic vacuum), and another beyond it
(the thermal bath). 
%

\section{$\rm \bf \Large \Rmnum{6}$ Environmental correlation function vs photon statistics}

In what follows we employ Eq.~(\ref{drhoNM}) (together with
Eqs.~(\ref{Mmic2},\ref{L0},\ref{Gamma})) for the investigation of
the thermal environmental correlation function using data of the
single molecule photon statistics. Several methods for counting
spontaneous photon emission events were proposed in the past.
Prominent examples are the quantum jumps approach \cite{QJ}, or the
generating function method \cite{ZB}, which combines analyticity,
calculational simplicity (for the low-dimensional systems) and
intuition. Even though originally developed for the Markovian
Optical Bloch Equations \cite{CT}, the generating function approach
may be quite easily generalized to the time-retarded equation of the
type of Eq.~(\ref{drhoNM}) \cite{Budini}. Setting $\!t_0\!=\!0$ and
transforming Eq.~(\ref{drhoNM}) to Laplace space $t\rightarrow\zeta$
we have
%
%
\begin{equation}
\vec{\rho}_S(\zeta)=\left[\zeta-\hat{{\cal L}}-{\hat{\cal
M}}_T(\zeta)-\hat{\Gamma}\right]^{-1}\vec{\rho}_S(0)\equiv\hat{\cal
G}(\zeta)\vec{\rho}_S(0).\label{rhocalG}
\end{equation}
The propagator $\hat{\cal G}(\zeta)$ may
be formally expressed in terms of the iterative expansion
$$
\!\hat{\cal G}(\zeta)\!= \left[1+\hat{\cal G}_0(\zeta)\hat{\Gamma}+
\hat{\cal G}_0(\zeta)\hat{\Gamma}\hat{\cal G}_0(\zeta)\hat{\Gamma}
+_{\cdots}\right]\hat{\cal G}_0(\zeta)=$$\vspace{-.5cm}\begin{equation}\vspace{-.5cm}\left[1-\hat{\cal
G}_0(\zeta)\hat{\Gamma}\right]^{-1}\!\hat{\cal G}_0(\zeta)=
\sum_{n=0}^\infty\hat{\cal G}^{(n)}(\zeta),
\label{Exp}
\end{equation}
\vspace*{-.1cm} where \vspace*{-.1cm}
\begin{equation}
\hat{\cal G}^{(n)}(\zeta)=\left(\hat{\cal
G}_0(\zeta)\hat{\Gamma}\right)^n\hat{\cal G}_0(\zeta)\label{Un},
\end{equation}
\vspace*{-.2cm} and \vspace*{-.2cm}
\begin{equation}
\hat{\cal G}_0(\zeta)= \left[\zeta-\hat{{\cal L}}-{\hat{\cal
M}}_T(\zeta)\right]^{-1} \label{U0}
\end{equation}
is the propagator of Eq.~(\ref{rhocalG}) without $\hat{\Gamma}$. The operator
$\hat{\Gamma}$, given by Eq.~(\ref{Gamma}), couples the population of the
excited state $\rho_{\rm ee}$ directly to the population $\rho_{\rm
gg}$ of the ground state, and hence, describes an incoherent
transition, which (within the first order perturbation approximation
with respect to the linear coupling to the radiation field) may be
associated with the spontaneous emission of a single photon
\cite{CT,Mandel,Mukamel}. With such an interpretation, each term
$\hat{\cal G}^{(n)}(\zeta)$ of Eq.~(\ref{Exp}) corresponds
to the two-level system evolution conditioned by $n$ spontaneous photon emission events.\\

The iterative expansion of Eq.~(\ref{Exp}) may be compactly
represented using the generating function \cite{ZB} \vspace*{-.3cm}
\begin{equation}
\vec{G}(t,s)=\sum_{n=0}^\infty s^n\vec{\rho}^{~(n)}(t), \label{G}
\end{equation}
where
\begin{equation}
\vec{\rho}^{(n)}(t)= \hat{{\cal
G}}^{(n)}(t,0)\vec{\rho}(0)\label{rhon},\end{equation}
and the parameter $s$ serves as an odometer of the spontaneous
emission events. Substituting the definition Eq.~(\ref{G}) into
Eq.~(\ref{rhocalG}) yields 
%
\begin{equation}
\vec{G}(\zeta,s)=\left[\zeta-\hat{{\cal L}}-{\hat{\cal
M}}_T(\zeta)-s\hat{\Gamma}\right]^{-1}\vec{G}(0,s),\\\label{NMGomega}
\end{equation}
which differs from Eq.~(\ref{rhocalG}) only by the extra factor $s$
multiplying $\hat{\Gamma}$. Since
\begin{equation}
\vec{\rho}^{~(n)}(t)=\frac{\partial^n \vec{G}(t,s)}{\partial
s^n}|_{s=0},\label{rhovsG}
\end{equation}
the statistics of the photon emission events may be fully obtained
by the derivatives of $\vec{G}(t,s)$ with respect to $s$. The most common
and simply measured quantities are the probability of $n$
spontaneous photon emissions
\begin{equation}
\tilde{P}_n(\zeta)\!=\!\tilde{\rho}_{\rm ee}^{(n)}(\zeta)+\tilde{\rho}_{\rm
gg}^{(n)}(\zeta)\!=\!\left[\frac{\partial^n}{\partial s^n}\tilde{G}_{\rm
ee}(s,\zeta)\!+\!\frac{\partial^n }{\partial s^n}\tilde{G}_{\rm
gg}(s,\zeta)\right]_{s=0}\!,\label{Pn}
\end{equation}
the mean photon number
\begin{equation}\langle \tilde{n}(\zeta)\rangle=\sum_{n=0}^\infty
n \tilde{P}_n(\zeta)=\left[\frac{\partial \tilde{G}_{\rm
ee}(s,\zeta)}{\partial s}+\frac{\partial \tilde{G}_{\rm
gg}(s,\zeta)}{\partial s}\right]_{s=1},\label{N}
\end{equation}
and the second moment
$$
\langle \tilde{n}^2(\zeta)\rangle=\left[\frac{\partial^2
\tilde{G}_{\rm ee}(s,\zeta)}{\partial s^2}+\frac{\partial^2
\tilde{G}_{\rm gg}(s,\zeta)}{\partial
s^2}\right]_{s=1}+$$\vspace{-.5cm}\begin{equation}+\left[\frac{\partial \tilde{G}_{\rm
ee}(s,\zeta)}{\partial s}+\frac{\partial \tilde{G}_{\rm
gg}(s,\zeta)}{\partial s}\right]_{s=1}\label{N2}.\\
\end{equation}
%

Using
Eqs.~(\ref{L0},\ref{Gamma},\ref{Mmic2},\ref{TABC},\ref{NMGomega},\ref{Pn},\ref{N},\ref{N2})
it is now straightforward to determine the connection between the
thermal noise correlation function $\beta_T(t)$ and single molecule
photon statistics. For this purpose we first need to find the Laplace
transform of the thermal memory kernel $\hat{\cal{M}}_T(t)$ Eqs.~(\ref{Mmic2},\ref{TABC}).
It follows from Eqs.~(\ref{TABC}) that the matrix elements of $\hat{\cal{M}}_T(t)$ may be
decomposed as
\begin{widetext}
\begin{equation}\begin{array}{c}
A_T(t)=i\frac{\Omega_0^2}{\Omega^2}\left[2{\rm Im}[\beta_T(t)]\frac{\Delta_L}{\Omega_0}+{\rm e}^{-it\Omega}
{\rm Im}[\beta_T(t)]\left(\frac{\Omega}{\Omega_0}-\frac{\Delta_L}{\Omega_0}\right)-
{\rm e}^{it\Omega}
{\rm Im}[\beta_T(t)]\left(\frac{\Omega}{\Omega_0}+\frac{\Delta_L}{\Omega_0}\right)\right],\\ \\
\!\!\!\!\!\!\!\!B_T(t)\!=\!-\frac{\Omega_0^2}{\Omega^2}\!\left[\!2{\rm Re}[\beta_T(t)]\!+\!{\rm e}^{-it\Omega}
{\rm Re}[\beta_T(t)]\!\left(\!1\!+\!2\frac{\Delta_L}{\Omega_0}\!\left(\!\frac{\Delta_L}{\Omega_0}\!-\!
\frac{\Omega}{\Omega_0}\!\right)\!\right)\!+\!
{\rm e}^{it\Omega}
{\rm Re}[\beta_T(t)]\!\left(\!1\!+\!2\frac{\Delta_L}{\Omega_0}\!\left(\!\frac{\Delta_L}{\Omega_0}\!+\!
\frac{\Omega}{\Omega_0}\!\right)\!\right)\!\right],\\ \\
C_T(t)=\frac{\Omega_0^2}{\Omega^2}\left[2{\rm Re}[\beta_T(t)]-{\rm e}^{-it\Omega}{\rm Re}[\beta_T(t)]-{\rm e}^{it\Omega}
{\rm Re}[\beta_T(t)]\right].\label{ABC}
\end{array}
\end{equation}
Since for a function of time $f(t)$, we have $L_{t\rightarrow \zeta}\left[{\rm e}^{\pm i\Omega t}
f(t)\right]=\tilde{f}(\zeta\mp i\Omega)$, the Laplace transform of Eqs.~(\ref{ABC}) yields
\begin{equation}\begin{array}{c}
\!\!\tilde{A}_T(\zeta)=i\frac{\Omega^2_0}{\Omega^2}\left[2\frac{\Delta_L}{\Omega_0}{\rm
Im}[\tilde{\beta}_T(\zeta)]+{\rm Im}[\tilde{\beta}_T(\zeta+i\Omega)](\frac{\Omega}{\Omega_0}-\frac{\Delta_L}{\Omega_0})-
{\rm
Im}[\tilde{\beta}_T(\zeta-i\Omega)](\frac{\Omega}{\Omega_0}+\frac{\Delta_L}{\Omega_0})
\right],\\ \\
\!\!\!\!\!\!\!\!\tilde{B}_T(\zeta)\!=\!-\frac{\Omega_0^2}{\Omega^2}\!\left[\!{2\rm
Re}[\tilde{\beta}_T(\zeta)]\!+\!
\!{\rm Re}[\tilde{\beta}_T\!(\!\zeta\!+i\Omega)\!]\left(\!1\!+\!2\frac{\Delta_L}{\Omega_0}\!\left(\!\frac{\Delta_L}{\Omega_0}\!-\!\frac{\Omega}{\Omega_0}\!\right)\!\right)
\!+\! {\rm
Re}[\tilde{\beta}_T(\!\zeta\!-i\Omega)]
\!\left(\!1\!+\!2\frac{\Delta_L}{\Omega_0}\!\left(\!\frac{\Delta_L}{\Omega_0}\!+\!
\frac{\Omega}{\Omega_0}\!\right)\!\right)\!\!\right],\\ \\
\!\tilde{C}_T(\zeta)\!=\!\frac{\Omega_0^2}{\Omega^2}\left[2{\rm
Re}[\tilde{\beta}_T(\zeta)]\!
\!-\!{\rm Re}[\tilde{\beta}_T(\zeta\!+\!i\Omega)]-\! {\rm
Re}[\tilde{\beta}_T(\zeta\!-\!i\Omega)]\right].\\
\end{array}
\label{LapABC}
\end{equation}

%
%
The explicit calculations following the prescription
Eqs.~(\ref{L0},\ref{Gamma},\ref{Mmic2},\ref{NMGomega},\ref{Pn},\ref{N},\ref{N2},\ref{LapABC})
may be done with the help of computational programs such as
Mathematica. To illustrate the method we confine the following
example to frequently used experimental conditions. Assuming that at
$t=0$ the emitter is prepared in the pure ground state and the laser
frequency is close to the resonance; expanding the
results in a Taylor series in $\Delta_L$ up to the first order we find 
$$
\tilde{G}_{\rm ee}(\zeta,s)+\tilde{G}_{\rm gg}(\zeta,s)\approx
\frac{\left(8\tilde{a}+\gamma+2\zeta\right)\left(\zeta+\gamma\right)+2\Omega_0^2}{\zeta\left(8\tilde{a}+\gamma+2\zeta\right)
\left(\zeta+\gamma\right)+\left(\gamma(1-s)+2\zeta\right)\Omega_0^2}+$$\vspace{-.5cm}\begin{equation}
\!\!+\!\Delta_L\frac{2(s\!-\!1)\gamma\left((\gamma\!+\!\zeta)(8\tilde{a}\!+\!\gamma\!+\!2\zeta)\!+\!2\Omega_0^2\right)
\left(2\tilde{b}(\gamma\!+\!2\zeta)\!+\!8\tilde{a}_+(\tilde{b}\!-\!\tilde{b}_-)\!-\!
(\gamma\!+\!2\zeta\!+\!2i\Omega_0)\tilde{b}_-\!+\!8\tilde{a}_-(\tilde{b}\!-\!\tilde{b}_+)\!-\!
(\gamma\!+\!2\zeta\!-\!2i\Omega_0)\tilde{b}_+\right)}{\left(\zeta(\gamma+\zeta)(8\tilde{a}+\gamma+2\zeta)+
(\gamma(1-s)+2\zeta)\Omega_0^2\right)^2\left(\gamma+2\zeta+4(\tilde{a}_-+\tilde{a}_+)\right)},
\label{LapABCres}
\end{equation}
\end{widetext}
%
%
where we used the shorthand notation: $\tilde{a}=\tilde{a}(\zeta)={\rm Re}[\tilde{\beta}(\zeta)]$,
$\tilde{b}=\tilde{b}(\zeta)={\rm Im}[\tilde{\beta}(\zeta)]$, $a_\pm=a_\pm(\zeta)={\rm Re}[\tilde\beta(\zeta\pm i\Omega)]$ and
$\tilde{b}_\pm=\tilde{b}_\pm(\zeta)={\rm Im}[\tilde\beta(\zeta\pm i\Omega)]$. Eq.~(\ref{LapABCres})
indicates that on resonance $\Delta_L=0$ the photon statistics
depends on $\tilde{a}={\rm Re}\left[\tilde{\beta}(\zeta)\right]$ alone, which
makes the reconstruction of the real part of the thermal correlation
function particularly simple.
Substituting Eq.~(\ref{LapABCres}) into
Eqs.~(\ref{Pn},\ref{N},\ref{N2}) in case of on resonance excitation
yields
%
%
$$
\!\!\!\!\tilde{P}_n(\zeta)=\frac{n!\gamma^n\Omega_0^{2n}\left((\gamma+\zeta)(\gamma+2\zeta+8\tilde{a})
+2\Omega_0^2\right)}
{\left((\gamma+2\zeta)(\gamma\zeta+\zeta^2+\Omega_0^2)+8\tilde{a}\zeta(\gamma+\zeta)\right)^{n+1}}\label{Pnvsa},
$$\vspace{-.4cm}$$
\langle \tilde{n}(\zeta)\rangle=\frac{\gamma\Omega_0^2}{\zeta^2\left((\gamma+\zeta)(\gamma+2\zeta+8\tilde{a})+
2\Omega_0^2\right)}\label{nvsa},
$$\vspace{-.4cm}
\begin{equation}
\langle \tilde{n}^2(\zeta)\rangle=\frac{\gamma\Omega_0^2(\gamma+\zeta)\left(\gamma\zeta+
2(\zeta^2+\Omega_0^2)+
8\tilde{a}\zeta\right)}{\zeta^3\left((\gamma+\zeta)(\gamma+2\zeta+8\tilde{a})+
2\Omega_0^2\right)^2}\label{n2vsa}.
\end{equation}
On specifying $J(\omega)$ and the temperature $T$, which according
to Eq.~(\ref{beta}) are needed for the calculation of the real part
of the thermal environmental correlation function ${\rm
Re}\left[\beta_T(t)\right]=\int_0^\infty d\omega
J(\omega)\coth\left[\frac{\omega}{2\kappa_B
T}\right]\cos\left[\omega t\right]$, it is straightforward to use
Eqs.~(\ref{n2vsa}) for predicting the corresponding photon
statistics. The converse procedure of reconstruction of the thermal
correlation function is obtained by inverting Eqs.~(\ref{n2vsa}).
For example, expressing $\tilde{a}={\rm Re}[\tilde{\beta}(\zeta)]$
in terms of $\tilde{P}_0(\zeta)$ and $\langle
\tilde{n}(\zeta)\rangle$, which are understood to be measured
experimentally, with the help of Eqs.~(\ref{n2vsa}) respectively
yields
\begin{equation}
\!\!{\rm Re}[\tilde{\beta}(\zeta)]\!=\!\frac{(\gamma\!+\!\zeta)(\gamma\!+\!2\zeta)\!+\!
2\Omega_0^2\!-\!(\gamma\!+\!2\zeta)(\zeta(\gamma\!+\!\zeta)\!+\!
\Omega_0^2)\tilde{P}_0(\zeta)}{8(\gamma+\zeta)(\zeta \tilde{P}_0(\zeta)-1)}
\label{avsP0},
\end{equation}
\begin{equation}
{\rm Re}[\tilde{\beta}(\zeta)]=\frac{\gamma\Omega_0^2-\langle \tilde{n}(\zeta)\rangle\zeta^2((\gamma+\zeta)(\gamma+2\zeta)+2\Omega_0^2)}{
8\langle \tilde{n}(\zeta)\rangle\zeta^2(\gamma+\zeta)}\label{avsn}.
\end{equation}
Note that Eqs.~(\ref{avsP0},\ref{avsn}) establish a well-defined
relation between
$\tilde{P}_0(\zeta)$ and $\langle \tilde{n}(\zeta)\rangle$. Expressing $\tilde{a}={\rm Re}[\tilde{\beta}(\zeta)]$ in terms of $\tilde{P}_n(\zeta)$ with $n>0$ and $\langle \tilde{n}^2(\zeta)\rangle$ is slightly more complicated and may entail multiple possibilities, which must be examined from
the physical point of view.
These expressions, which we shall skip since they are massive,
may be easily obtained with the help of Mathematica if needed.\\

\begin{figure}[t] 
\begin{tabular}{cc}
\hspace{-.3cm}\includegraphics[scale=0.35]{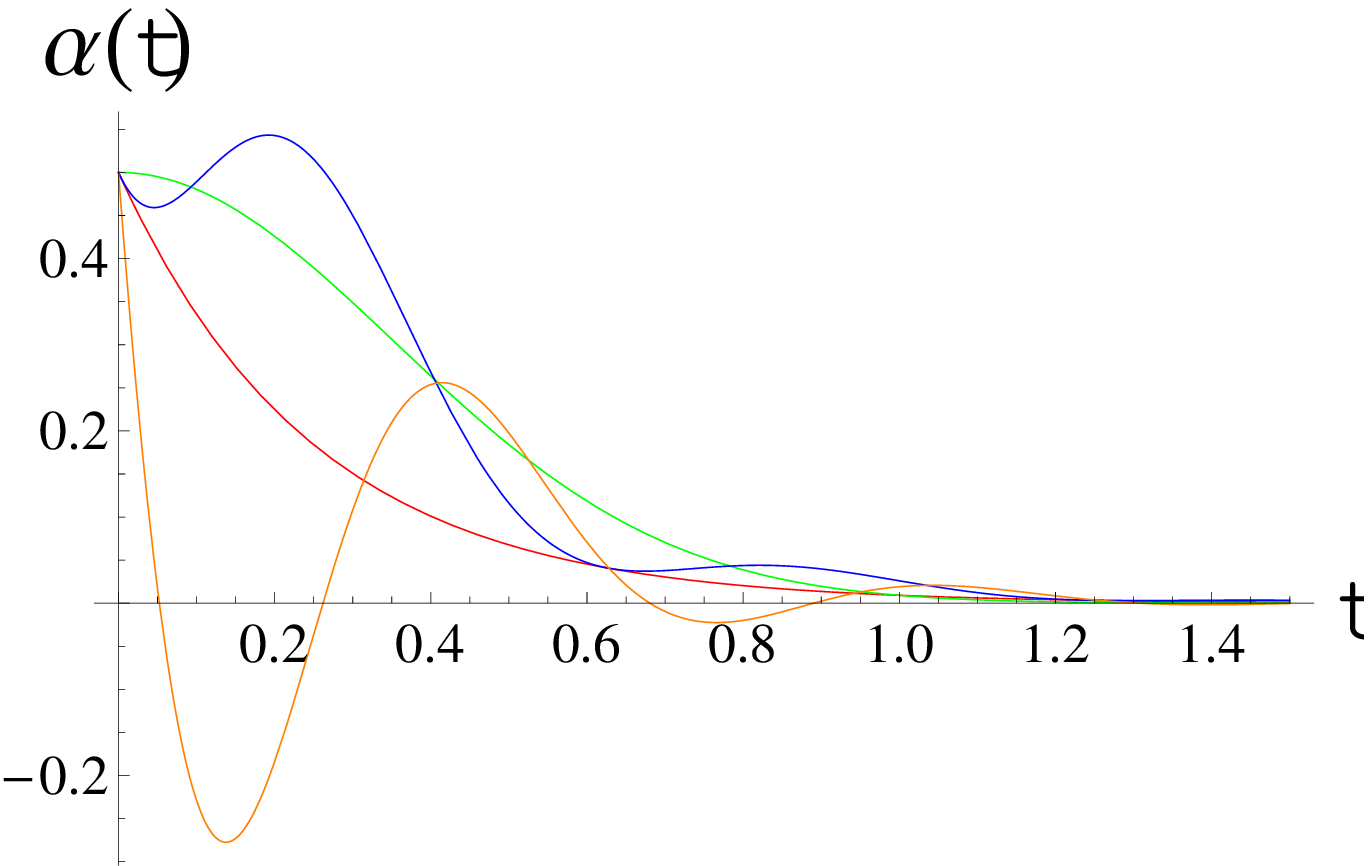}&\hspace{-.15cm}\includegraphics[scale=0.35]{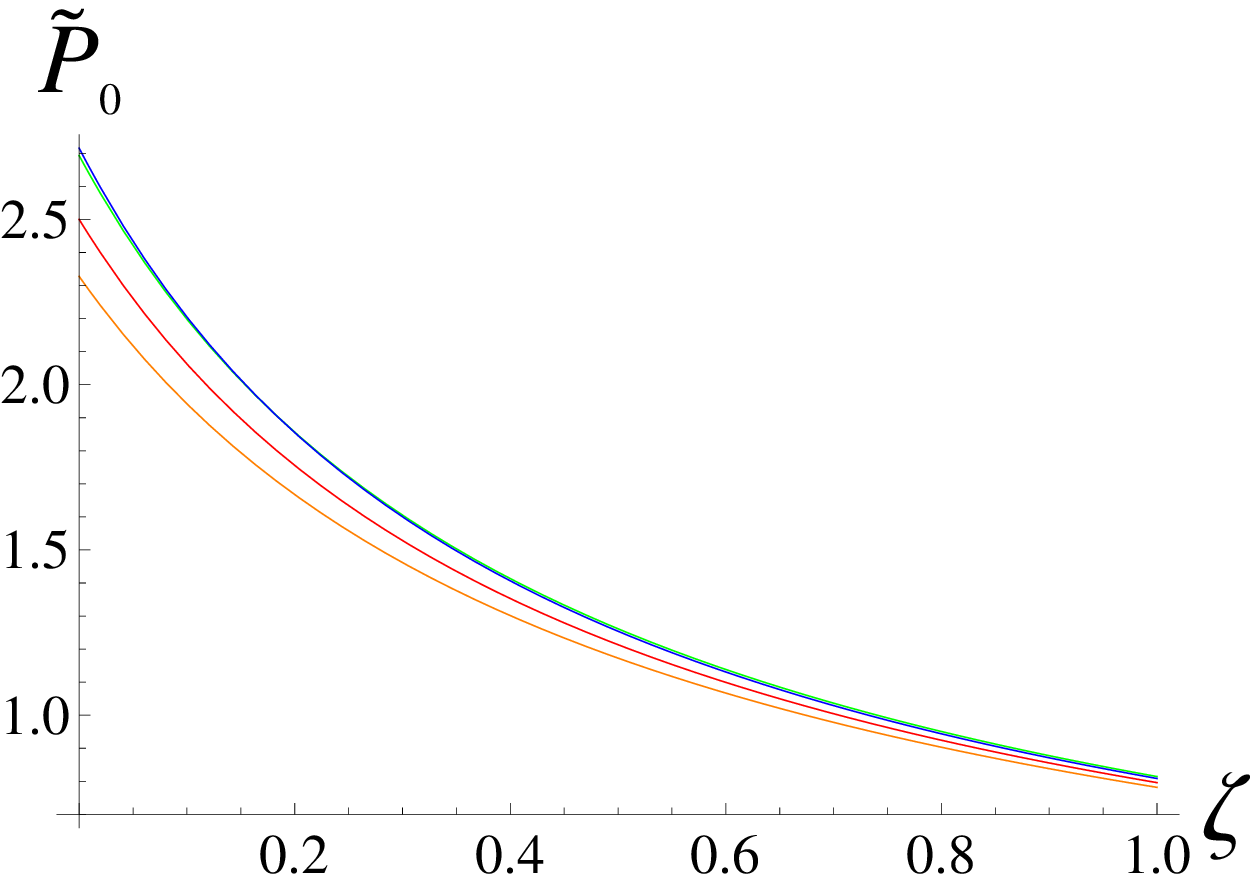}\\
\newline
$(a)$&
$(b)$
\end{tabular}
\caption{\label{P0}(Color online)(a) The plot of $\beta(t)=A{\rm e}^{-Rt};~A{\rm e}^{-Rt^2};~(1-A\cos[Bt]){\rm e}^{-Rt};~
(A-\sin[Bt]){\rm e}^{-Rt}$, for $A=0.5$ and $B=10$, represented respectively
by the red, blue, green, and orange curves;(b) The plot of $\tilde{P}_0(\zeta)$ (Eq.~(\ref{n2vsa})), for $\gamma=1,\Omega_0=2$,
resulting from a different choice of the environmental
correlation function in Fig.~2(b). The colors of the curves are matched .}
%
\end{figure}

Finally, we would like to illustrate the results of
Eqs.~(\ref{n2vsa}). To do so, using Eq.~(\ref{beta}), we must choose
an explicit form of the environmental spectral function $J(\omega)$.
The latter, if microscopically unknown, may be modeled
phenomenologically, for example, as a power-law
$J(\omega)=\eta_s\omega^s\omega_c^{1-s}{\rm e}^{-\omega/\omega_c}$,
where $\eta_s$ is the viscosity coefficient, $\omega_c$ is a cutoff
frequency and $s$, for the case of the spin-boson model, is the
dimension of space \cite{Weiss}. However, the integral of
Eq.~(\ref{beta}) induced by such a choice of spectral function, is
not analytically solvable in parametric form.
\renewcommand{\baselinestretch}{1}
\begin{figure}[t]
\centering
\includegraphics[scale=0.5]
{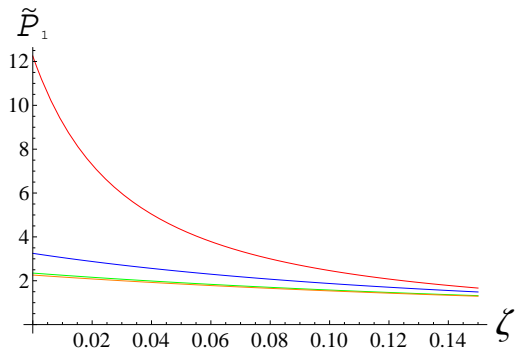}
\caption{\label{P1}(Color online) The plot of $\tilde{P}_1(\zeta)$ (Eq.~(\ref{n2vsa})) for $\gamma=1,\Omega_0=2$,
corresponding to environmental
correlation function $\beta(t)=\nu^2{\rm e}^{-Rt}$ for $\nu=1$ and $R=0.1,1,10,100$ represented respectively
by the red, blue, green, and orange curves.}%
\end{figure}
Hence, we restrict the illustration of Eqs.~(\ref{n2vsa}) to
some arbitrary examples of the environmental correlation function.
In Fig.~\ref{P1} we plot the probability of one photon emission, setting $\beta(t)=\nu^2{\rm e}^{-R t}$,
which is a standard assumption of the spectral diffusion approach \cite{PRL,JCP}. The graph, comparing $\tilde{P}_1(\zeta)$
for $R=0.1,1,10,100$ (red, blue, green, orange) shows that an increase in the order of $R$
requires nearly exponential improvement of the measurement accuracy.
In Fig.~\ref{P0} we examine the dependence of the probability of zero photon emission events $\tilde{P}_0(\zeta)$ (Fig.~\ref{P0}(b))
on several shapes of $\beta(t)$ (Fig.~\ref{P0}(a)), chosen such that the autocorrelation time $\tau_c$ and the maximal value
of $\beta(t)$ are close. The fact that for different choices of $\beta(t)$ we arrive at effectively the same result for $\tilde{P}_0(\zeta)$
may be seen as a justification of the phenomenological assumption $\beta(t)=\nu^2{\rm e}^{-R t}$,
used to plot Fig.~\ref{P1}.

\section{$\rm \bf \Large \Rmnum{7}$ Summary}
\vspace{-.2cm} By superposing an approximate phenomenological Dyson
equation with a microscopic reduced master equation, we provided a
correspondence between the microscopic and the spectral diffusion
approaches. It was shown that excluding the interaction of the
two-level system emitter with the laser, the dynamics governed by
the spectral diffusion model is indistinguishable from the dynamics
obtained by the microscopic theory, whenever the evolution of the
environment is not determined on the microscopic level. This allowed
to identify, up to a constant, the spectral diffusion correlation
function with the real part of the thermal environmental correlation
function. Furthermore, it was demonstrated that the real problem of
using the spectral diffusion method for the description of standard
SMS experiments beyond the Markovian limit, is the independent rates
of variation assumption, which can be overcome using the microscopic
reduced propagator theory \cite{Vega,DGS,TYu}. Finally, combining
the methods of the latter with the generalized generating functions
approach, we have established analytic expressions for the thermal
correlation function allowing to extract information on the
environmental noise beyond the Markovian approximation from
the measured single emitter photon statistics.\\
%

\section*{Appendix A: Derivation of Dyson equation}
\setcounter{equation}{0}
\renewcommand{\theequation}{A.\arabic{equation}}

In this appendix we provide the details of derivation of Eqs.~(\ref{GDE},\ref{calM}).
Substituting $\vec{\rho}(t|\eta_t)=\hat{U}(t,t_0|\eta_t)\vec{\rho}(t_0)$, where $\hat{U}(t,t_0|\eta_t)$
is the propagator, into Eq.~(\ref{SchEq}) we have
\begin{equation}
\frac{d}{dt}\hat{U}(t,t_0|\eta_t)=\
\hat{\cal O}\left[\hat{U}(t,t_0|\eta_t),t\right]+\eta_t\hat{\Upsilon}\hat{U}(t,t_0|\eta_t),\label{ASchEq}
\end{equation}
and rewrite the latter equation in the integral form \cite{CT}
\begin{equation}
\hat{U}(t,t_0|\eta_t)=
\hat{U}_0(t,t_0)+\int_{t_0}^t\hat{U}_0(t,\tau)\eta_\tau\hat{\Upsilon}\hat{U}(t,t_0|\eta_t)d\tau,\label{ISchEq}
\end{equation}
where $\hat{U}_0(t,t_0)$ is the solution of Eq.~(\ref{GL0E}). 
Further, we expand Eq.~(\ref{ISchEq})
by iteration
$$
\!\hat{U}(t,t_0|\eta_t)=\hat{U}_0(t,t_0)+\int_{t_0}^t dt_1\hat{U}_0(t,t_1)\eta_{t_1}\hat{\Upsilon}\hat{U}_0(t_1,t_0)+$$\vspace{-.5cm}$$+
\int_{t_0}^t dt_1\int_{t_0}^{t_2} dt_1\hat{U}_0(t,t_2)\eta_{t_2}\hat{\Upsilon}\hat{U}_0(t_2,t_1)\eta_{t_1}\hat{\Upsilon}\hat{U}_0(t_1,t_0)+$$\vspace{-.5cm}$$
\!\!\!\!+
\!\int_{t_0}^t\!\!dt_3\!\!\int_{t_0}^{t_3}\! \!\!dt_2\!\!\int_{t_0}^{t_2}\!\! \!\!\!\!dt_1\hat{U}_0(t,t_3)\eta_{t_3}\hat{\Upsilon}\hat{U}_0(t_3,t_2)\eta_{t_2}\hat{\Upsilon}\times
$$\vspace{-.5cm}$$\times\hat{U}_0(t_2,t_1)\eta_{t_1}\hat{\Upsilon}\hat{U}_0(t_1,t_0)+$$\vspace{-.5cm}$$
\!\!+\!\int_{t_0}^t\!\!\!dt_4\!\!\int_{t_0}^{t_4}\!\!\!dt_3\!\!\int_{t_0}^{t_3}\!\!\!dt_2\!\!\int_{t_0}^{t_2}\!\!\!dt_1\hat{U}_0(t,t_4)\eta_{t_4}\hat{\Upsilon}\hat{U}_0(t_4,t_3)\times$$
\vspace{-.5cm}\begin{equation}\times\eta_{t_3}\hat{\Upsilon}\hat{U}_0(t_3,t_2)\eta_{t_2}\hat{\Upsilon}\hat{U}_0(t_2,t_1)\eta_{t_1}\hat{\Upsilon}\hat{U}_0(t_1,t_0)+... ~.\label{A2}
\end{equation}
Assuming $\eta_t$ is a zero mean Gaussian noise, we take an average of Eq.~(\ref{A2}). Discarding the
long time correlations contribution, as discussed in Section $\rm \Rmnum{3}$, this gives
$$
\!\hat{U}(t,t_0)=\hat{U}_0(t,t_0)+$$\vspace{-.6cm}$$+
\int_{t_0}^t dt_1\int_{t_0}^{t_2} dt_1\langle\eta_{t_2}\eta_{t_1}\rangle\hat{U}_0(t,t_2)\hat{\Upsilon}\hat{U}_0(t_2,t_1)\hat{\Upsilon}\hat{U}_0(t_1,t_0)+$$\vspace{-.5cm}$$
\!\!+\!\int_{t_0}^t\!\!\!dt_4\!\!\int_{t_0}^{t_4}\!\!\!dt_3\!\!\int_{t_0}^{t_3}\!\!\!dt_2\!\!\int_{t_0}^{t_2}\!\!\!dt_1
\langle\eta_{t_4}\eta_{t_3}\rangle\langle\eta_{t_2}\eta_{t_1}\rangle\hat{U}_0(t,t_4)\times$$
\vspace{-.5cm}\begin{equation}\times\hat{\Upsilon}\hat{U}_0(t_4,t_3)\hat{\Upsilon}\hat{U}_0(t_3,t_2)\hat{\Upsilon}\hat{U}_0(t_2,t_1)\hat{\Upsilon}\hat{U}_0(t_1,t_0)+...,
\end{equation}
where $\hat{U}(t,t_0)=\langle \hat{U}(t,t_0|\eta_t)\rangle$. Evidently, the above iterative expansion is equivalent to
$$
\!\hat{U}(t,t_0)=\hat{U}_0(t,t_0)+$$\vspace{-.6cm}\begin{equation}+
\int_{t_0}^t dt_1\int_{t_0}^{t_2} dt_1\langle\eta_{t_2}\eta_{t_1}\rangle\hat{U}_0(t,t_2)\hat{\Upsilon}\hat{U}_0(t_2,t_1)\hat{\Upsilon}\hat{U}(t_1,t_0),
\end{equation}
which in turn, by analogy with
Eqs.~(\ref{ASchEq},\ref{ISchEq}), is equivalent to the differential equation
\begin{equation}
\!\!\frac{d}{dt}\hat{U}(t,t_0)\!=\!\hat{\cal O}\left[\hat{U}(t,t_0),t\right]\!+\!
\int_{t_0}^t \!\!d\tau\langle\eta_{t}\eta_{\tau}\rangle\hat{\Upsilon}\hat{U}_0(t,t_\tau)\hat{\Upsilon}\hat{U}(t_\tau,t_0).
\end{equation}
Finally, acting with the last equation on the initial state $\vec{\rho}(t_0)$ yields Eqs.~(\ref{GDE},\ref{calM}).

\section*{Appendix B: Reduced propagators approach with two reservoirs}
\setcounter{equation}{0}
\renewcommand{\theequation}{B.\arabic{equation}}

In this appendix we consider a closed system, composed of a
two-level particle interacting with a thermal reservoir, a
monochromatic laser field and the electromagnetic vacuum. The total Hamiltonian within the rotating
wave approximation (RWA) with respect to the coupling to the electromagnetic field is \cite{CT, Scully,Vega,DGS,TYu}
\begin{equation}
\begin{array}{c}
\hat{H}(t)=\frac{\omega_0}{2}\hat{\sigma}_z+ (\hat{\sigma}_+
\frac{\Omega_0}{2}{\rm e}^{-i\omega_Lt}+\hat{\sigma}_-
\frac{\Omega_0}{2}{\rm e}^{i\omega_Lt})+\\\\+\sum_\lambda\omega_\lambda
\hat{b}^{\dag}_\lambda \hat{b}_\lambda+ \sum_\mu\omega_\mu \hat{a}^{\dag}_\mu \hat{a}_\mu
+\\
\\+\sum_\lambda g_\lambda\hat{\sigma}_z(\hat{b}^{\dag}_\lambda+\hat{b}_\lambda)+\sum_\mu p_\mu(\hat{\sigma}_+ \hat{a}_\mu+\hat{\sigma}_- \hat{a}^{\dag}_\mu),\end{array}\label{4H}
\end{equation}
where $\hat{b}^{\dag}_{\lambda},\hat{b}_{\lambda}$ and
$\hat{a}^{\dag}_{\lambda},\hat{a}_{\lambda}$ are the boson ladder operators of
the thermal environment and the electromagnetic field respectively, the Pauli
matrices $\hat{\sigma}_i$ represent the particle operators, and $\Omega_0$ is the Rabi
frequency of the laser pump oscillating at frequency $\omega_L$.
Transforming to the rotating frame by
\begin{equation}
\hat{R}(t)=\exp\left[i\frac{\omega_L}{2}t\hat{\sigma}_z+\sum_\mu
i\omega_Lt\hat{a}^{\dag}_\mu \hat{a}_\mu\right],\label{Rot}
\end{equation}
we suppress the time dependence in
Eq.~(\ref{4H}), which yields Eq.~(\ref{Hrwa}) of the article.\\

Given Eq.~(\ref{Hrwa}), extending
the methods of \cite{Vega,DGS,TYu} for simultaneous interaction of the
particle with two bosonic reservoirs, we consider the total
propagator of the system in the partial representation picture with
respect to $\hat{H}_B+\hat{H}_R=\sum_{\lambda}\omega_{\lambda}\hat{b}^{\dag}_{\lambda}\hat{b}_{\lambda}+
\sum_{\mu}(\omega_{\mu}-\omega_L)\hat{a}^{\dag}_{\mu}\hat{a}_{\mu}$:
\begin{equation}
\hat{U}_I(t,{t_0})={\rm e}^{i(\hat{H}_B+\hat{H}_R)(t-t_0)}\hat{U}(t-{t_0}){\rm
e}^{-i(\hat{H}_B+\hat{H}_R)(t-t_0)}, \label{UI}
\end{equation}
and define the reduced propagator
\begin{equation}
\hat{G}(\vec{y}_t^*\vec{z}_t^*\vec{y}_{t_0}\vec{z}_{t_0}|t
t_0)\equiv\langle
\vec{y}_t\vec{z}_t|\hat{U}_I(t,{t_0})|\vec{y}_{t_0}\vec{z}_{t_0}\rangle,\label{Gzy}
\end{equation}
where $|\vec{z}_t\rangle=\prod_\lambda |z_{t,\lambda}\rangle$
describes the thermal field in the Bargmann coherent
states representation (and similarly, $|\vec{y}_t\rangle$ represents the state of the electromagnetic field). 
Applying $\hat{G}(\vec{y}_t^*\vec{z}_t^*\vec{y}_{t_0}\vec{z}_{t_0}|t
t_0)$ to the initial state of the two-level system, propagates the
latter such that the final state is simultaneously conditioned by
specific evolution trajectories of both reservoirs. In what follows
we show that under certain circumstances, the particle conditional
density matrix
$\hat{\rho}_I(\vec{y}_t^*\vec{z}_t^*\vec{y}_{t_0}\vec{z}_{t_0}|t
t_0)$ depends on the environmental degrees of freedom through the
stochastic processes $y_t,z_{t}$ and their autocorrelation functions
$\alpha(t),\beta(t)$, defined below. This constitutes a
simplification of the general case, where higher order
cross-correlation functions of
$y_t$ and $z_{t}$ can be involved as well. \\

The time evolution equation for
$\hat{G}(\vec{y}_t^*\vec{z}_t^*\vec{y}_{t_0}\vec{z}_{t_0}|t t_0)$ is
obtained from the projection of the Schr\"{o}dinger equation for
$\hat{U}_I(t,t_0)$, given by Eq.~(\ref{UI}):
\begin{equation}
\begin{array}{c}
\frac{\partial}{\partial
t}\hat{G}(\vec{y}_t^*\vec{z}_t^*\vec{y}_{t_0}\vec{z}_{t_0}|t
t_0)=\langle \vec{y}_t\vec{z}_t|\frac{\partial}{\partial
t}\hat{U}_I(t,{t_0})|\vec{y}_{t_0}\vec{z}_{t_0}\rangle
=\\\\=-i\left[\hat{H}_S+\hat{\sigma}_z\sum_\lambda g_\lambda{\rm
e}^{i\omega_\lambda
t }z_{t,\lambda}^*+\right.\\ \\
\left.+ \hat{\sigma}_-\sum_\mu p_\mu{\rm
e}^{i(\omega_\mu-\omega_L) t} y_{t,\mu}^*\right]\hat{G}(\vec{y}_t^*\vec{z}_t^*\vec{y}_{t_0}\vec{z}_{t_0}|t t_0)-\\\\
- i\hat{\sigma}_z\sum_\lambda g_\lambda{\rm e}^{-i\omega_\lambda
t}\langle \vec{y}_{t} \vec{z}_{t}|\hat{b}_\lambda
\hat{U}_I(t,{t_0})|\vec{y}_{{t_0}}\vec{z}_{{t_0}}\rangle -
\\\\-i\hat{\sigma}_+\sum_\mu p_\mu{\rm e}^{-i(\omega_\mu-\omega_L)
t}\langle \vec{y}_{t}\vec{z}_{t}|\hat{a}_\mu
\hat{U}_I(t,{t_0})|\vec{y}_{t_0}\vec{z}_{t_0}\rangle,
\end{array}
\label{DtGzy}
\end{equation}
where the last two terms constitute an obstacle for getting a closed
equation for
$\hat{G}(\vec{y}_t^*\vec{z}_t^*\vec{y}_{t_0}\vec{z}_{t_0}|t t_0)$.
We use the same scheme as above to represent these terms as
functions of
$\hat{G}(\vec{y}_t^*\vec{z}_t^*\vec{y}_{t_0}\vec{z}_{t_0}|t t_0)$.
Starting with the thermal reservoir we have
\begin{equation}
\begin{array}{c}
\langle\vec{y}_{t}\vec{z}_{t}|\hat{b}_\lambda
\hat{U}_I(t,{t_0})|\vec{y}_{t_0}\vec{z}_{t_0}\rangle=\\\\=\langle
\vec{y}_{t}\vec{z}_{t}| \hat{U}_I(t,{t_0})\hat{U}_I^{-1}(t,{t_0})
\hat{b}_\lambda
\hat{U}_I(t,{t_0})|\vec{y}_{t_0}\vec{z}_{t_0}\rangle =\\
\\= \langle \vec{y}_{t}\vec{z}_{t}|
\hat{U}_I(t,{t_0})\hat{b}_\lambda(t,
{t_0})|\vec{y}_{t_0}\vec{z}_{t_0}\rangle,\end{array}
\end{equation}
where
\begin{equation}
\hat{b}_\lambda(t, {t_0})\equiv \hat{U}_I^{-1}(t,{t_0})\hat{b}_\lambda
\hat{U}_I(t,{t_0}).\label{bfi}
\end{equation}
Rewriting Eq.~(\ref{bfi}) in the integral form yields
\begin{equation}
\hat{b}_\lambda(t,
{t_0})=\hat{b}_\lambda-ig_\lambda\int_{t_0}^{t}\hat{\sigma}_z(\tau,{t_0}){\rm
e}^{i\omega_\lambda\tau} d\tau,\label{Eqbfi}
\end{equation}
where $\hat{\sigma}_z(\tau,{t_0})=\hat{U}_I^{-1}(t,{t_0})\hat{\sigma}_z
\hat{U}_I(t,{t_0})$.
%
%
Repeating the same procedure with respect to the radiation field and
inserting the results back into Eq.~(\ref{DtGzy}) gives
\begin{equation}
\begin{array}{c}
\frac{\partial}{\partial
t}\hat{G}(\vec{y}_t^*\vec{z}_t^*\vec{y}_{t_0}\vec{z}_{t_0}|t t_0)=
\left[-i\hat{H}_S+\hat{\sigma}_zz_{t}^*-\hat{\sigma}_z
z_{t_0}+\right.\\\\\left.+
\hat{\sigma}_-y_t^*-\hat{\sigma}_+ y_{t_0}\right]\hat{G}(\vec{y}_t^*\vec{z}_t^*\vec{y}_{t_0}\vec{z}_{t_0}|t t_0)-\\
\\- i\hat{\sigma}_z\int_{t_0}^t\beta(t-\tau)\langle \vec{y}_{t}\vec{z}_{t}|
\hat{U}_I(t,t_0)\hat{\sigma}_z(\tau,t_0)|\vec{y}_{t_0}\vec{z}_{t_0}\rangle
d\tau -\\ \\- i\hat{\sigma}_+\int_{t_0}^t\alpha(t-\tau)\langle
\vec{y}_{t}\vec{z}_{t}|\hat{U}_I(t,t_0)\hat{\sigma}_-(\tau,t_0)|\vec{y}_{t_0}\vec{z}_{t_0}\rangle
d\tau,
\end{array}
\label{dGzy}
\end{equation}
where
\begin{equation}
z_{\tau}\equiv-i\sum_\lambda g_\lambda{\rm e}^{i\omega_\lambda t}
z_{\tau,\lambda}, ~y_\tau\equiv-i\sum_\mu p_\mu{\rm
e}^{i(\omega_\mu-\omega_L) t} y_{\tau,\mu},\label{noises}
\end{equation}
and
\begin{equation}
\beta(t)\equiv\sum_\lambda |g_\lambda|^2{\rm e}^{-i\omega_\lambda t},~\alpha(t)\equiv \sum_\mu |p_\mu|^2{\rm e}^{-i(\omega_\mu-\omega_L) t}.\\
\label{cfs}
\end{equation}

Further, in order to close Eq.~(\ref{dGzy}), we are interested in rearranging the last two terms in Eq.~(\ref{dGzy}) as
\begin{equation}\begin{array}{c}
\!\!\!\!\langle
\vec{y}_{t}\vec{z}_{t}|\hat{U}_I(t,{t_0})\hat{\sigma}_i(\tau,{t_0})|\vec{y}_{t_0}\vec{z}_{t_0}\rangle\!=\!
\langle
\vec{y}_{t}\vec{z}_{t}|\hat{\sigma}_i(\tau,t)\hat{U}_I(t,{t_0})|\vec{y}_{t_0}\vec{z}_{t_0}\rangle\\
\\ =
\hat{O}(y^*_tz^*_ty_{t_0}z_{t_0}
t\tau)\hat{G}(\vec{y}^*_t\vec{z}_t^*\vec{y}_{t_0}\vec{z}_{t_0}|t
{t_0}),
\end{array}\label{COG}
\end{equation}
where $\hat{O}(y^*_tz^*_ty_{t_0}z_{t_0} t\tau)$ is an operator
acting in the particle subspace. This can be done by formally
integrating and iterating the Heisenberg equation for
$\hat{\sigma}_i(\tau,t)$ in power series of $g_\lambda$  and
$p_\mu$. Taking into account that the autocorrelation functions
$\alpha(t),\beta(t)$ are already of the second order in $g_\lambda$
and $p_\mu$ respectively, keeping only the zeroth order solution for
$\hat{\sigma}_i(\tau,t)$:
\begin{equation}
\hat{\sigma}_{i}(\tau,t)={\rm e}^{-iH_S(t-\tau)}\hat{\sigma}_{i}{\rm
e}^{iH_S(t-\tau)}+{\cal O}(g) \label{Expsigma}
\end{equation}
accounts for a satisfactory approximation in the case $g_\lambda$
and $p_\mu$ are small. Substituting the truncated expansion
Eq.~(\ref{Expsigma}) back into Eq.~(\ref{dGzy}) we neglect all the
terms proportional to $g_\lambda^np_\mu^m$ with $m+n>2$. Since the
zeroth order approximation Eq.~(\ref{Expsigma}) is restricted to the
particle subspace, we can pull it out of the brackets in
Eq.~(\ref{COG}), which allows to close the equation for
$\hat{G}(\vec{y}^*_t\vec{z}_t^*\vec{y}_{t_0}\vec{z}_{t_0}|t {t_0})$
and yields
\begin{equation}
\begin{array}{c}
\frac{\partial}{\partial t}\hat{G}(\vec{y}^*_t\vec{z}_t^*\vec{y}_{t_0}\vec{z}_{t_0}|t {t_0})=\\\\
=\left[-i\hat{H}_S+\hat{\sigma}_zz_{t}^*-\hat{\sigma}_z z_{t_0}+\hat{\sigma}_-y_t^*-\hat{\sigma}_+ y_{t_0}\right.-\\
\\- i\hat{\sigma}_z\int_{t_0}^t\beta(t-\tau){\rm
e}^{-i\hat{H}_S(t-\tau)}\hat{\sigma}_{z}{\rm
e}^{i\hat{H}_S(t-\tau)}-\\ \\\left.-
i\hat{\sigma}_+\int_{t_0}^t\alpha(t-\tau){\rm
e}^{-i\hat{H}_S(t-\tau)}\hat{\sigma}_{-}{\rm
e}^{i\hat{H}_S(t-\tau)}\right]\times\\\\ \times
\hat{G}(\vec{y}^*_t\vec{z}_t^*\vec{y}_{t_0}\vec{z}_{t_0}|t {t_0})
d\tau.
\end{array}
\label{DGzy}\\
\end{equation}
The lesson of Eq.~(\ref{DGzy}) is that within the second order
approximation in the interaction magnitudes, the effective coupling
between the two reservoirs is eliminated.
The reduced evolution of the particle is given by direct
independent summation of the contribution arising from the interaction with
each one of the fields alone.\\

From now on the results of \cite{Vega,DGS,TYu} may be adopted directly. Using
Eq.~(\ref{DGzy}) the master equation for the conditional density matrix follows from
\begin{equation}\begin{array}{c}
\frac{\partial}{\partial
t}\hat{\rho}_I(\vec{y}^*_t\vec{z}^*_t\vec{y}_{t_0}\vec{z}_{t_0}|t
t_0)=\frac{\partial}{\partial t} \left[\hat{G}(\vec{y}^*_t\vec{z}_t^*\vec{y}_{t_0}\vec{z}_{t_0}|t {t_0})\times\right.\\\\
\left.\times
\hat{\rho}_I(\vec{y}^*_{t_0}\vec{z}^*_{t_0}\vec{y}_{t_0}\vec{z}_{t_0}|t_0
t_0)\hat{G}(\vec{y}^*_t\vec{z}_t^*\vec{y}_{t_0}\vec{z}_{t_0}|t {t_0})\right].
\end{array}
\label{DtrhoS}
\end{equation}
Afterwards, the evolution equation for the marginal density matrix $\hat{\rho}_S(t)$ is obtained by integrating out
the irrelevant environmental degrees of freedom, taking into account
the initial Boltzmann equilibrium distribution of the thermal bath and the zero temperature $\delta$-correlated
distribution of the electromagnetic vacuum. This procedure, requiring
an application of the generalized Novikov theorem,
was worked out in \cite{Vega,DGS,TYu}. Using the results for each of the two
reservoirs respectively we obtain Eq.~(\ref{dtroConv}). \\

\end{document}